\documentclass[a4paper,11pt]{article}

\pdfoutput=1 

\usepackage{jcappub} 

\usepackage[T1]{fontenc} 

\usepackage{hyperref}
\usepackage[dvipsnames]{xcolor}

\usepackage{tikz}
\usepackage{tikz-feynman}

{}

\newcommand{\beq}{\begin{equation}}  
\newcommand{\eeq}{\end{equation}}  
\newcommand{\beqa}{\begin{eqnarray}}  
\newcommand{\eeqa}{\end{eqnarray}}  
\newcommand{\bea}{\begin{eqnarray}}  
\newcommand{\eea}{\end{eqnarray}}

\newcommand{\MDM}{M_{\rm DM}}
\newcommand{\tDM}{\tau_{\rm DM}}

\newcommand{\PhiG}{\phi_{\rm G}}
\newcommand{\PhiEG}{\phi_{\rm EG}}

\newcommand{\psiGC}{\psi_{\rm GC}}
\newcommand{\PDM}{P_{\rm DM}}
\newcommand{\Vis}{\mathcal{V}}
\newcommand{\TSi}{\mathrm{TS}_i}
\newcommand{\TStot}{\mathrm{TS}_{\rm tot}}

\newcommand{\pval}{p\text{-value}}
\newcommand{\km}{{\rm\,km}}
\newcommand{\kpc}{{\rm\,kpc}}
\newcommand{\GeV}{{\rm\,GeV}}

\newcommand{\EeV}{{\rm\,EeV}}

\newcommand{\kmt}{{\rm KM3-230213A}}

\graphicspath{{figures/}}

\usepackage{orcidlink}

\title{Earth rotation turns event timing into a geometric probe of UHE neutrino origin}

\author[a,b]{Andrew Cheek\,\orcidlink{0000-0002-8773-831X},}

\author[a,b]{João Paulo Pinheiro\,\orcidlink{0000-0002-6536-2040},}

\author[c,d]{and Jordi~Salvado\,\orcidlink{0000-0002-7847-2142}}

\affiliation[a]{Tsung-Dao Lee Institute \& School of Physics and Astronomy, Shanghai Jiao Tong
University, Shanghai 200240, China}

\affiliation[b]{Key Laboratory for Particle Astrophysics and Cosmology (MOE) \& Shanghai Key Laboratory for Particle Physics and Cosmology, Shanghai Jiao Tong University, Shanghai 200240, China}

\affiliation[c]{Departament de Física Quàntica i Astrofísica (FQA), Universitat de Barcelona (UB), c. Martí i Franqués, 1, 08028 Barcelona, Spain}

\affiliation[d]{Institut de Ciències del Cosmos (ICCUB), Universitat de Barcelona (UB), c. Martí i Franqués, 1, 08028 Barcelona, Spain}

\abstract{
It has been proposed that the ultra-high-energy (UHE) event 
$\rm KM3-230213A$ detected by KM3NeT could be explained by dark matter 
(DM) decay. Prima facie this seems unlikely because the arrival direction 
of the event is opposite to the Galactic Centre. We develop a per-event 
test statistic to quantitively assess this possibility and forecast the 
required future events to exclude the DM hypothesis in favour of an 
isotropic signal. For the single event observed, the DM decay hypothesis 
is disfavoured but not excluded ($\pval_{\rm DM}\simeq0.13$--$0.15$). We 
emphasise how including the time-averaged detector visibility helps 
discrimination despite reducing the proportion of visible sky, reducing 
the number of events for exclusion from $\sim33$--$43$ to $\sim22$--$27$. 
Moving beyond this, we perform a fully time-resolved forecast and find 
that the required number of events for exclusion reduces by $40\%$, 
$\sim14$--$16$. 
The time variation in the signal provides vital information allowing one 
to exclude or confirm the DM hypothesis with much fewer events.  
Our results are robust against DM decay channels and halo distributions 
and can readily be applied other relics distributed similarly. Our 
framework allows one to turn event timing into a probe of signal geometry 
and is generic to any UHE equatorial neutrino telescopes.  
}

\begin{document}

\pagestyle{myplain}

\maketitle

\flushbottom

\section{Introduction}

In its first year of data taking, the neutrino telescope KM3NeT detected the  ultra-high energy (UHE) event, $\kmt$~\cite{KM3NeT:2025npi}, reconstructed at a peak of $\sim 220\,{\rm PeV}$-- the most energetic neutrino ever observed. Its inferred single-flavor flux is in mild tension with the non-observation of comparable events at IceCube~\cite{IceCube:2016zyt}, which has ignited an intense interpretative effort.

Speculation on the origin of $\kmt$ is predictably
comprehensive. Astrophysical explanations include individual
point sources~\cite{KM3NeT:2025bxl,Dzhatdoev:2025sdi,Wang:2025lgn, Yuan:2025zwe}, cosmogenic neutrinos from UHE 
cosmic-rays~\cite{KM3NeT:2025vut, Zhang:2025abk,Boxi:2025ony}, as well as primordial neutrinos~\cite{GrimbaumYamamoto:2026kam}. The $\gamma$--ray counterpart expected from such explanations has also been explored~\cite{Fang:2025nzg,Filipovic:2025ulm,Crnogorcevic:2025vou,Das:2025vqd}. More exotic possibilities include evaporating primordial black holes~\cite{Dvali:2025ktz,Klipfel:2025jql,Boccia:2025hpm,Choi:2025hqt,Airoldi:2025opo,Airoldi:2025bgr} to new neutrino interactions~\cite{Brdar:2025azm,Neronov:2025jfj,Alves:2025xul}, Lorentz-invariance violation~\cite{KM3NeT:2025mfl,Cattaneo:2025uxk,Yang:2025kfr,Bertolez-Martinez:2025trs}, and super-heavy decaying dark matter (DM)~\cite{Barman:2025hoz,Jho:2025gaf,Borah:2025igh,Kohri:2025bsn,Narita:2025udw,Khan:2025gxs,Jiang:2025blz}. In this paper, we study the spatial distribution of events from DM decay. Our analysis is purely spatial, it directly applies to any hypothesis that predicts an anisotropy correlated with the DM density and can be extended to other anisotropic signal models. 

The difficulty in interpreting $\kmt$ as the product of a DM decay, is the 
arrival direction, $(\mathrm{RA},\mathrm{Dec})=(94.3^\circ,-7.8^\circ)$, 
almost \emph{precisely opposite} the direction of the Galactic Center 
(GC). This is at odds with expectation because the neutrino flux from 
Galactic DM decay correlates with the density of the halo and is therefore 
strongly peaked towards the GC. An initial DM-induced signal arriving from 
the anti-centre is, at first thought, improbable. But is it impossible? 
Our goals are to quantify \emph{how} improbable it is, to determine 
whether the decaying DM hypothesis can already be excluded as an 
explanation, and if not, to determine how many additional UHE events 
KM3NeT must collect to confirm or exclude it. We note that some have 
already tried to address the problem of $\kmt$ directionality by 
introducing particle models where decays only occur at higher 
redshift~\cite{Jho:2025gaf}, such scenarios may be constrained by 
cosmological probes~\cite{Hambye:2021moy}, but here we try to determine 
whether it is strictly necessary.  

In the last decade a similar story played out for the events detected by IceCube~\cite{IceCube:2013cdw,IceCube:2013low, IceCube:2014stg, Cholis:2012kq,Anchordoqui:2013dnh, Murase:2014tsa}. 
The possibility that such high-energy neutrinos came from dark matter was explored in Refs.~\cite{Feldstein:2013kka,Murase:2015gea,Chianese:2016smc,Cohen:2016uyg}, 
but after leveraging the direction information of events and geometry of 
the dark matter profile, it was found that only a subset of
events could originate from dark matter~\cite{Bai:2013nga}.
We build on the framework of Ref.~\cite{Bai:2013nga}. The role of detector 
visibility in the $\kmt$ direction has been noted before through the time 
and azimuthal-\emph{averaged} fractional exposure ~\cite{Aloisio:2025nts}.
The time information has been exploited to test a transient PBH-
evaporation origin of the event, where the time-dependent exposure of 
$\gamma$-ray and neutrino
observatories is essential to predict the multi-messenger signal in the 
hours preceding the burst~\cite{Airoldi:2025opo}.

In this article, our novel contribution is to show that, even for the 
(quasi-)stationary signal models such as the anisotropic flux of DM decay, 
including the \emph{time-resolved} information has substantial impact on 
discrimination power. This is a consequence of the equatorial positioning 
of KM3NeT, where the instantaneous rotating field of view, makes the 
arrival time of each event an independent observable. This temporal 
information, discarded by any
time-averaged or all-sky analysis, is the central tool of this work.
The remainder of the article develops the DM flux model 
(Sec.~\ref{sec:flux}), the visibility and
time-dependence machinery (Sec.~\ref{sec:vis}), the test statistic
(Sec.~\ref{sec:ts}), and the results (Sec.~\ref{sec:results}).

\section{Neutrino Sky Maps from Dark Matter Decay}
\label{sec:flux}

The differential neutrino flux from dark matter decay can be separated into two
components, the Galactic component and the extra-galactic (isotropic) component, \begin{equation}
\frac{d^2\phi_{\nu}}{dEd\Omega}= \frac{d^2\phi_\nu^{\rm gal.}}{dEd\Omega} + \frac{d^2\phi_\nu^{\rm EG}}{dEd\Omega}\,.
\end{equation}

The Galactic flux is the line-of-sight integral of the halo density
$\rho_{\rm DM}$ towards $(b,l)$,
\begin{equation}
\frac{d^2\phi_{i,\rm G}(E_i)}{dE_i d\Omega} = \frac{1}{4\pi M_{\rm DM} \tau_{\rm DM}} \frac{dN_i(E_i)}{dE_i}\times \int_0^{s_{\rm max}} ds\,\rho_{\rm DM}(r(s,b,l))
\end{equation}
with the decay spectrum $dN/dE$, the DM mass $\MDM$ and lifetime $\tDM$. The
angular dependence is conveniently captured by the $\mathcal{D}$-factor,
\begin{equation}
\mathcal{D} = \frac{1}{\Delta\Omega} \int_{\Delta\Omega} d\Omega \int_0^{s_{\rm max}} ds \, \rho_{\rm DM}(s,b,l)\,.
\end{equation}
Where the galactocentric radius along a line of sight is:
\begin{equation}
r_{GC}(s, l, b) = \sqrt{s^2 + r_\odot^2 - 2 s r_\odot \cos b \cos l}
\end{equation}
with $s$ being the distance from the observer, and $(l,b)$ are galactic
coordinates.  In these coordinates the KM3NeT direction is 
$(l,b)=(209.4^\circ,-10.7^\circ)$, i.e. nearly anti-GC.

For the halo model, a standard choice is the Navarro--Frenk--White (NFW) profile,
\begin{equation}
\rho_{\rm NFW}(r) = \frac{\rho_s}{(r/r_s)(1 + r/r_s)^2}\,.
\end{equation}
with $r_s=25\kpc$ and $\rho_s=0.23\GeV/\mathrm{cm}^3$ as in
Ref.~\cite{KM3NeT:2026flu}. The profile uncertainty directly rescales the
relative DM flux in a given direction. In this work we account for this uncertainty by taking the range of inner Milky-Way profiles
compiled in Ref.~\cite{Hussein:2025xwm}. The profiles come from hydrodynamic simulations of
Milky-Way analogues---Auriga L3~\cite{Grand:2024xnm},
VINTERGATAN-GM~\cite{Rey:2022mwh}, TNG50~\cite{2024MNRAS.535.1721P}, and
FIRE-2~\cite{Wetzel:2022man}. Astrometric
determinations of the DM profile~\cite{Zhou:2022lar,2024MNRAS.528..693O} lie within this range. We label the two
extremes \textbf{restrictive} ($\rho_{\rm res}$) and \textbf{permissive} ($\rho_{\rm per}$) reflecting how constraining the respective profiles are. The centrally concentrated $\rho_{\rm res}$ is well described by adiabatic contraction~\cite{Gnedin:2004cx,Blumenthal:1985qy}, whereas the
shallower $\rho_{\rm per}$ is consistent with systems calculated with strong baryonic feedback~\cite{Wetzel:2022man}.
The restrictive profile enhances the GC peak and therefore yields the strongest
directional discrimination, as we confirm below.

Integrating over the cosmological DM density gives the isotropic flux
\begin{equation}
\frac{d\phi_{\rm EG}}{dEd\Omega} = \frac{\rho_{\rm DM}}{4\pi \MDM \tDM}
\int_0^{z_{\rm max}} dz\;\left|\frac{dt}{dz}\right|
\frac{dN}{dE'}\bigg|_{E' = E(1+z)},
\label{eq:phi_EG_sec}
\end{equation}
with $\rho_{\rm DM}=\Omega_{\rm DM}\rho_c$ the mean DM density
($\Omega_{\rm DM}=0.27$, $\rho_c=4.7\times10^{-6}\GeV/\mathrm{cm}^3$),
$H_0=67.3\,\mathrm{km\,s^{-1}\,Mpc^{-1}}$, $\Omega_m=0.315$,
$\Omega_\Lambda=0.685$, and kinematic cut-off
$z_{\rm max}^{\rm kin}=\MDM/(2E)-1$. The redshifting of the spectrum smooths any
sharp spectral features of the Galactic flux signal, these would be present for the primary $\nu\bar{\nu}$ final states for example.

Both components share the per-decay spectrum, which carries a monochromatic
\emph{primary} line (when a neutrino is a direct decay product) and a continuous
\emph{secondary} tail of final-state radiation,
\begin{equation}
\frac{dN}{dE} = \frac{dN_{\text{prim}}}{dE} + \frac{dN_{\text{sec}}}{dE}\,.
\end{equation}
We compute this with \texttt{HDMSpectra}~\cite{Bauer:2020jay}, which 
includes electroweak corrections validated for high-energies. Following 
the KM3NeT analysis~\cite{KM3NeT:2026flu} we take three benchmark channels 
with a $100\%$ branching ratio, $\nu\bar\nu$, $\tau^+\tau^-$ and $b\bar b$, at their best-fit masses
$\MDM=[0.2,\,1,\,100]\EeV$ and lifetimes
$\tDM=[4,\,7,\,4]\times10^{26}\,$s respectively. The three channels differ 
mainly in the whether they have primary component and where the end-point 
of the spectrum is at high energies. Later we will discuss how the Earth 
opacity and hence KM3NeT visibility is affected by neutrino energy, which 
in turn effects the results of our analysis. For example, the hardness of 
the $b\bar b$ spectrum, extends almost two decades beyond the $\nu\bar\nu$ 
and $\tau^+\tau^-$ cut-offs, will prove to be relevant for the discussion of Sec.~\ref{sec:results}.

\begin{figure}
\centering
\includegraphics[width=0.6\linewidth]{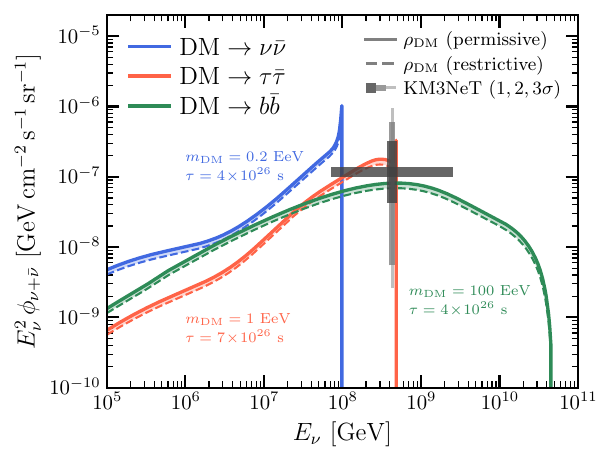}
\caption{$E_\nu^2$-weighted neutrino-plus-antineutrino flux from super-heavy DM decay in the direction of $\kmt$, for the three best-fit benchmarks of Ref.~\cite{KM3NeT:2026flu}: $\nu\bar\nu$ at $\MDM=0.2\EeV$, $\tDM=4\times10^{26}$\,s (blue); $\tau^+\tau^-$ at $1\EeV$, $7\times10^{26}$\,s (red); $b\bar b$ at $100\EeV$, $4\times10^{26}$\,s (green). Full (dashed) curves use the permissive (restrictive) halo profile. Grey bands show the KM3NeT $1,2,3\sigma$ flux intervals.}
\label{fig:fluxes_bmp}
\end{figure}

Figure~\ref{fig:fluxes_bmp} shows the resulting $E_\nu^2$-weighted flux in the
direction of $\kmt$, for the two profile extremes, compared with the KM3NeT
flux bands. The colored curves are the best-fit benchmarks of
Ref.~\cite{KM3NeT:2026flu}; the spread between the full (permissive) and dashed
(restrictive) lines illustrate the halo-profile uncertainty that propagates
into the test statistic and which is pronounced in this direction but more in towards the GC. 
A purely directional interpretation of $\kmt$ implicitly treats every 
point of the sky as equally observable. At the energy of the event this is 
badly violated: the neutrino nucleon cross-section is large enough that 
the Earth becomes opaque, and the detector is sensitive only to events 
arriving from a narrow range of directions near the local horizon.

\section{Detector visibility and arrival-time information}
\label{sec:vis}

An idealized directional interpretation of $\kmt$ implicitly treats every point of
the sky as equally observable. At the energy of the event this is badly
violated: the neutrino nucleon cross-section is large enough that the 
Earth becomes opaque, and the detector is sensitive only to events 
arriving from a narrow range of directions near the local 
horizon~\cite{Schumacher:2021mw,Schumacher:2025qca,Arguelles:2025ewg}. 
Because that range rotates with the Earth, the fraction of the halo that 
KM3NeT can see changes over the course of a day. Folding this time-
dependent coverage into the analysis is what turns the arrival
time of an event into directional information about its origin. We develop 
a simplified version of the machinery
here and collect the underlying chord geometry, Earth model and cross-
section parametrizations in
App.~\ref{app:visibility}. Note that Ref.~\cite{Palmisano:2026sid} 
recently put forward a semi-analytical framework for converting flux into 
event rates for neutrino telescopes.

We define the visibility $\Vis(E,\hat n,t)$ as the probability that a neutrino of
energy $E$ arriving from direction $\hat n$ at time $t$ produces a detectable
up-going lepton. KM3NeT/ARCA sits in the Mediterranean Sea at
\begin{equation}
(\mathrm{lat},\mathrm{lon})=(36.27^\circ\,\mathrm{N},\;16.10^\circ\,\mathrm{E}),
\qquad \text{depth } 3500\,\mathrm{m},
\label{eq:km3_location}
\end{equation}
which we treat as a single point detector. A neutrino reaching this point with zenith
angle $\theta_z$ has crossed a chord of the Earth; only the final stretch nearest
the detector, of length $L_{\rm eff}$ set by the secondary-muon range, can yield a
lepton that reaches the instrumented volume, while the remainder of the chord attenuates the flux. Writing the column depths of these two segments as
$X_{\rm screen}$ (attenuating) and $X_{\rm target}$ (interacting), the visibility
factorises into survival through the screen and interaction in the target,
\begin{equation}
\Vis(E,\hat n,t) =
\underbrace{\exp\!\left(-\frac{X_{\rm screen}\,\sigma_{\rm tot}(E)}{m_N}\right)}_{\text{survival}}
\times
\underbrace{\left[1 - \exp\!\left(-\frac{X_{\rm target}\,\sigma_{\rm CC}(E)}{m_N}\right)\right]}_{\text{interaction}}\,,
\label{eq:visibility}
\end{equation}
with $m_N=1.674\times10^{-24}\,\mathrm{g}$ the nucleon mass and
$\sigma_{\rm tot},\sigma_{\rm CC}$ the total and charged-current neutrino--nucleon
cross sections. The chord geometry that fixes $X_{\rm screen}$ and $X_{\rm target}$
as functions of $(\hat n,t)$, the muon range that sets $L_{\rm eff}$, the PREM
density profile~\cite{Dziewonski:1981xy} and the cross-section parametrisations are given in
App.~\ref{app:visibility}. In this article we do not apply the analysis cuts for the down-going events to reject the atmospheric muon background
performed by the KM3NeT collaboration and we do not add background sources
for our analysis.

Two limits of Eq.~\eqref{eq:visibility} are important for the analysis. 
Directions pointing deep below the horizon traverse the dense
inner core, where $X_{\rm screen}$ is so large that the survival factor is
exponentially suppressed and $\Vis\to0$. Detectable events are confined
to a band of directions just below the horizon in which is long enough a 
column of atoms 
to provoke an interaction, short enough to survive attenuation. The 
angular width of this band
shrinks as the energy, and hence the opacity, grows: this is the origin of 
the channel dependence found in Sec.~\ref{sec:results}, where harder decay 
spectra reach higher energies, narrow the band, and deepen the daily 
modulation that the timing exploits.

For a near-equatorial detector the visible band is not fixed on the sky: as the
Earth rotates, the band sweeps through all right ascensions once per day.
The arrival time of an event therefore selects which strip of the sky lay below
the horizon and was therefore observable at that instant. Concretely, the local
zenith angle of a fixed equatorial direction $(\delta,\alpha)$ is a function of
time through the hour angle (App.~\ref{app:time}), so $\Vis$ at fixed
$(\delta,\alpha)$ oscillates over the day. From KM3NeT the Galactic Centre
($\delta\approx-29^\circ$) transits the local meridian once per day,
reaching a maximum elevation of only $\sim25^\circ$ and spending only part of the
day inside the visible band. An event recorded while the GC lies within the band
is far stronger evidence for a DM origin than an otherwise identical event recorded
while the GC is hidden, since the two carry different information even when they arrive
from the same direction. This is precisely the observable that a time-averaged or
all-sky analysis discards.

\begin{figure}[t]
\centering
\includegraphics[width=0.6\linewidth]{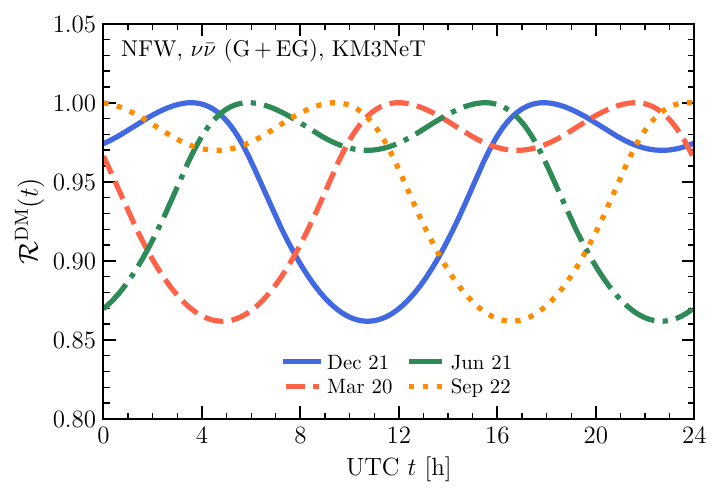}
\caption{Normalised DM-induced rate $\mathcal{R}^{\rm DM}(t)$ over a day
for the NFW profile and the $\nu\bar\nu$ channel (Galactic plus extragalactic),
at four epochs: Dec~21 (blue), Mar~20 (red dashed), Jun~21 (green dot-dashed),
Sep~22 (orange dotted). The daily peak corresponds to the GC transiting the
KM3NeT visible band.}
\label{fig:rate_vis}
\end{figure}

We quantify the effect through the (un-normalised) DM rate
$R(t)\equiv\int dE\,d\Omega\,\Vis\,\phi_{\rm DM}$ and its normalised form
\beq
\mathcal{R}^{\rm DM}(t)\equiv\frac{R(t)}{R^{\rm max}}\,,
\eeq
shown in Fig.~\ref{fig:rate_vis} at four times over a year. 
Note that the absolute normalisation of $R(t)$ is irrelevant for the
analysis: $\PDM$ in Sec.~\ref{sec:ts} is normalised by construction,
so the unknown factor $1/(\MDM\tDM)$ cancels and only the
shape of $\mathcal{R}^{\rm DM}(t)$ enters the test.
The signal is a daily modulation by around $15\%$ for the NFW-plus-extragalactic flux in the
$\nu\bar\nu$ channel with a peak when the GC transits the visible band. Over the year, there is a
seasonal drift in phase, reflecting the offset between solar and sidereal time. It is this modulation, not merely the daily-averaged exposure, that the
test statistic of Sec.~\ref{sec:ts} is built to exploit.

\section{Test statistic}
\label{sec:ts}

The event distribution is naturally expressed in equatorial coordinates, so we
map the DM flux from $(b,l)$ to $(\delta,\alpha)$ and define the normalised,
time-dependent DM probability density
\beq
\PDM(t,\delta,\alpha)=\frac{1}{\Phi_\nu}\frac{d\Phi_\nu(t,\delta,\alpha)}{d\Omega}\,,
\qquad
\Phi_\nu(t,\delta,\alpha)=\int dE_\nu\,\Vis(t,E_\nu,\delta,\alpha)\,\phi_{\rm tot}(E_\nu,\delta,\alpha)\,,
\eeq
with $\phi_{\rm tot}=\PhiG+\PhiEG$.
The probability density is evaluated on a $360\times180$ grid in $(\alpha,\delta)$
($1^\circ\times1^\circ$) and $280$ bins in time ($\simeq5$~min), 
with the flux integrated numerically over the remaining variables.

\begin{figure}[h!]
\centering
\includegraphics[width=0.95\linewidth]{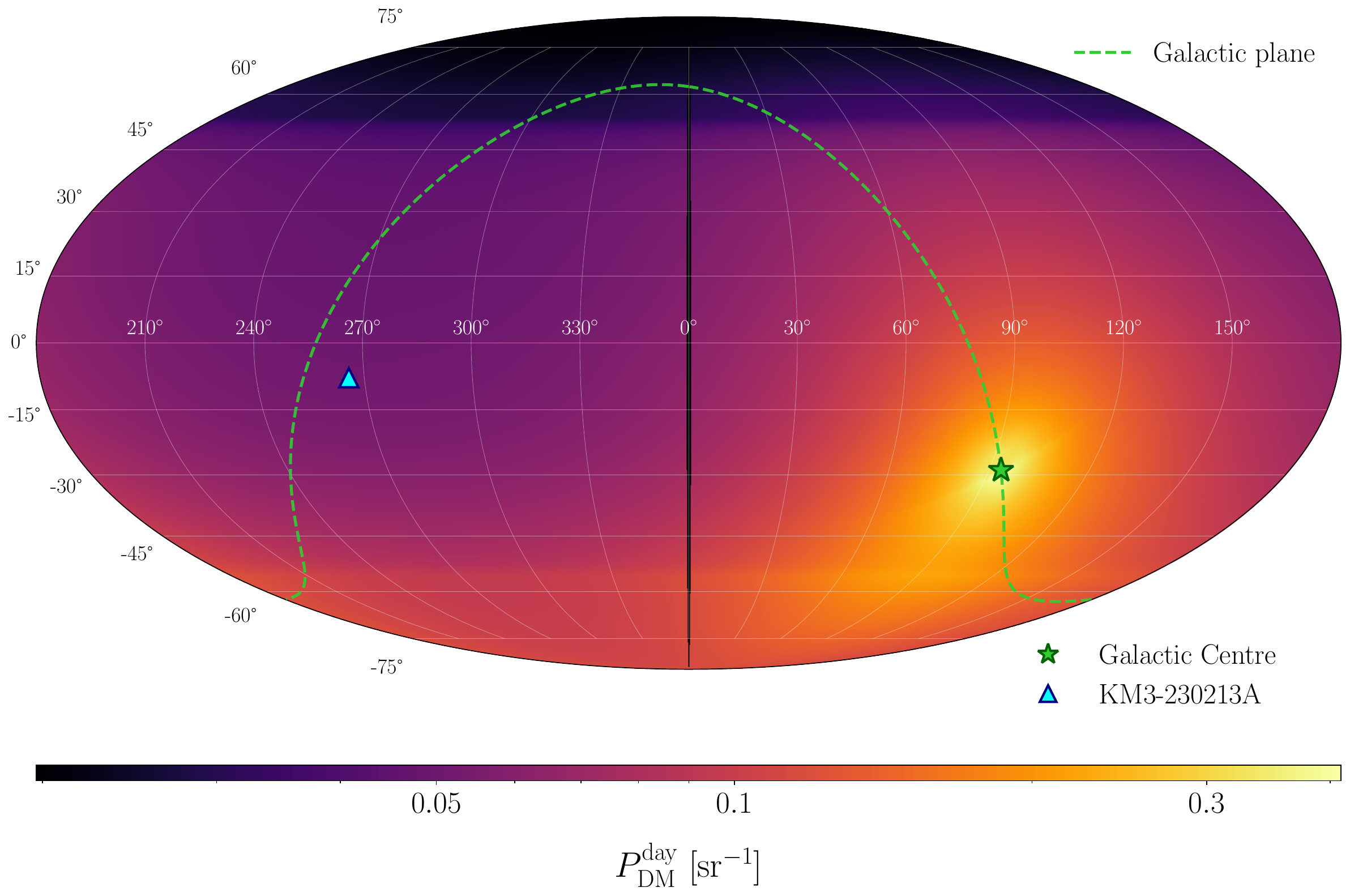}
\includegraphics[width=0.9\linewidth]{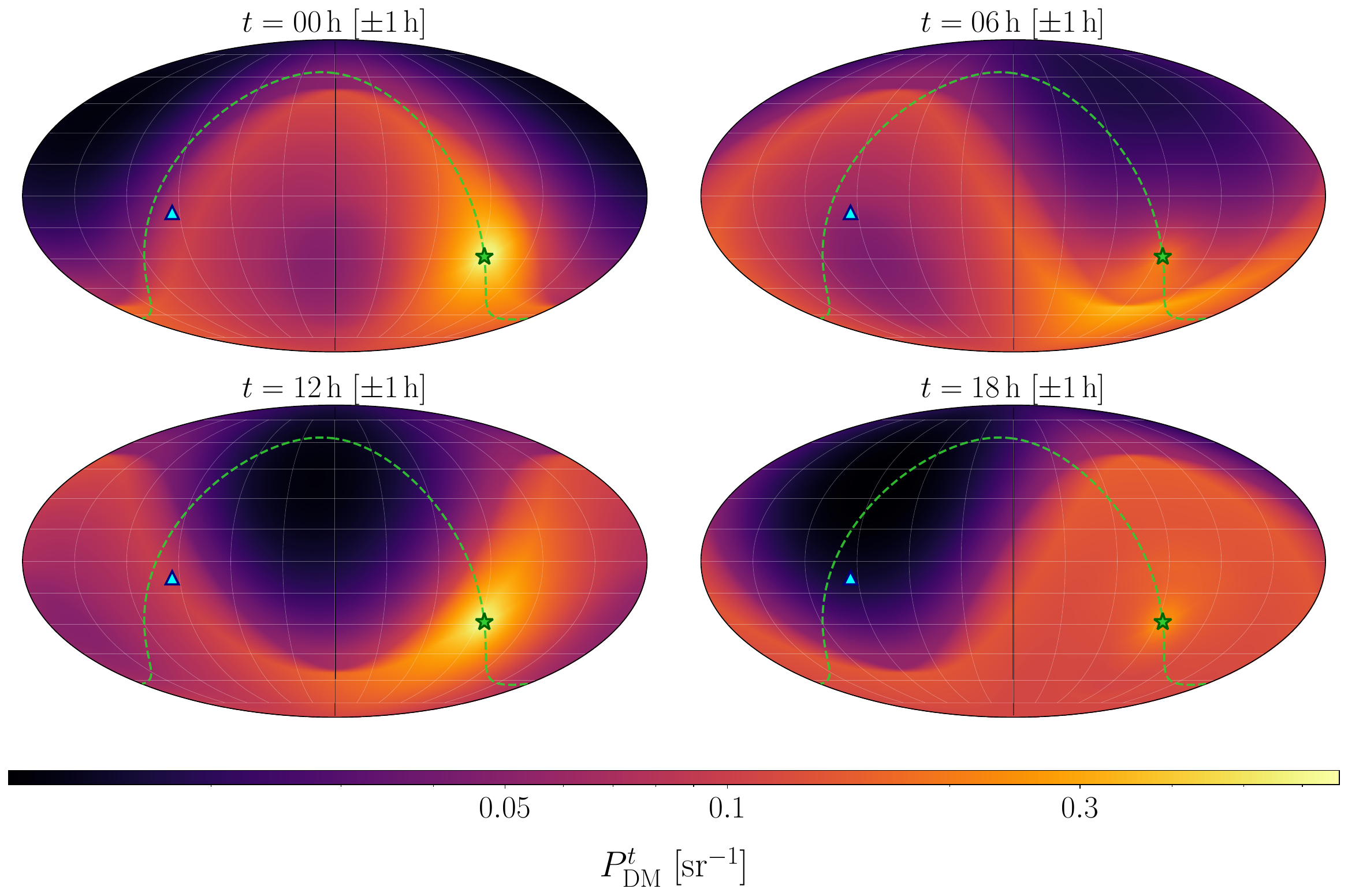}
\caption{  Mollweide projections of the DM probability map
$\PDM(\alpha,\delta)$ in equatorial coordinates for the NFW profile and $\nu\bar\nu$
channel.  Green star: Galactic Centre
$(\alpha,\delta)\approx(266.4^\circ,-28.9^\circ)$;
green dashed curve: Galactic plane; cyan triangle:
$\kmt$ at $(94.3^\circ,-7.8^\circ)$.
\textbf{Top:} day-averaged map $\PDM^{\rm day}(\alpha,\delta)$.
\textbf{Bottom:} instantaneous maps $\PDM^{t}(\alpha,\delta)$
at  hours $t=0,\,6,\,12,\,18$\,h
(each $\pm1$\,h band), sharing a common colour scale.%
}
\label{fig:pdm_skymap}
\end{figure}

Figure~\ref{fig:pdm_skymap} shows the probability density sky-map
$\PDM(\alpha,\delta)$ for the NFW profile and the $\nu\bar\nu$
channel, both day-averaged (top) and at four different times
$t=0,6,12,18$\,h with $\pm1$\,h bands (bottom), where $t=0$ is the time of $\kmt$.
The day-averaged panel makes visible the two effects discussed in
Sec.~\ref{sec:vis}: the peak in $\PDM(\alpha,\delta)$ aligning with the GC (green star) and
monotonic decrease of $\PDM(\alpha,\delta)$ towards the anti-centre, modulated by the
declination-dependent KM3NeT exposure that vanishes in the
circumpolar cap and is largest in the south. $\kmt$ (cyan
triangle) sits at $\psiGC\approx112^\circ$, deep in the
low-$\PDM$ anti-centre region; the left--right asymmetry reflects
the off-centre position of the Sun in the halo.

The four bottom panels show what time averaging hides. At a fixed instant the
visible region is a thin band near the local horizon which, although broad in
right ascension, sweeps across the sky as the Earth rotates. What localises
$\PDM=\Vis\,\phi_{\rm tot}/\Phi_\nu$ is not the width of this band but its overlap
with the GC-peaked flux. At $t=0$\,h the GC lies within the band, so the probability
density is dominated by the flux peak and concentrates around the GC
($\alpha\approx266^\circ$); at $t=6$ and $18$\,h the band has rotated by
$\pm90^\circ$ in RA, moving its overlap with the halo, so the peak of $\PDM$ shifts
but remains present; at $t=12$\,h the GC has rotated deep below the horizon, towards
the nadir where the Earth is opaque, leaving the visible band, and the most probable
accessible direction is no longer correlated with the galactic plane.

At the KM3NeT latitude the up-going requirement alone already removes a fraction
of visibility: sources with declination $\delta\gtrsim+54^\circ=90^\circ-36.3^\circ$ never
set below the local horizon and are permanently invisible, producing the 
northern dark region of the probability maps in Fig.~\ref{fig:pdm_skymap}. Towards the
south the fraction of the day during which a source is observable grows
monotonically, so the time-averaged exposure is already strongly
declination-dependent before any per-event timing is included.

For a given event at $(t_i, \delta_i, \alpha_i)$, we define the per-event statistic
\begin{equation}
\TSi = -\log P_{\text{DM}}(t_i, \delta_i, \alpha_i)
   \times \psi_{\text{GC}}(\delta_i, \alpha_i)\,,
\label{eq:ts_def_1ev}
\end{equation}
where $\psi_{\text{GC}}$ is the angular distance to the GC. For $N$ events the statistics is just the sum,
\beq
\TStot=\sum_{i=1}^N\TSi\,.
\label{eq:ts_total}
\eeq

With this definition, $\TStot$ quantifies the compatibility of a given measurement with the hypothesis, where larger values correspond to less compatible realisations. To relate the value measured in the observation, $\TStot^{\rm obs}$, to a $\pval$, we do not assume a known form for the test-statistic distribution. Instead, we generate Monte-Carlo realisations under the null hypothesis and study the resulting fluctuations of $\TStot$. The $\pval$ is then computed as the fraction of pseudo-experiments that yield $\TStot>\TStot^{\rm obs}$. Thus, the $\pval$ is the probability (under the null hypothesis) to obtain a test statistic at least as large as the observed one. A small $\pval$ disfavours the null hypothesis.

Since we rely on simulations, the specific definition of $\TSi$ is to some extent arbitrary, however in principle, alternative definitions may provide stronger discrimination. For a single event, because $\TSi$ is the product of two functions that both increase monotonically with distance from the Galactic Centre, including or omitting the angular factor does not change the ordering of realisations relative to $\TStot^{\rm obs}$, and the resulting $\pval$ is unchanged. This does not generally hold for multiple events, as shown in App.~\ref{app:ts_consistency}, the unweighted statistic $-\log\PDM$ achieves marginally stronger discrimination, so the thresholds reported below using Eq.~\eqref{eq:ts_def_1ev} are slightly conservative.
We retain the $\psiGC$ factor for two practical reasons: it regularises the $\PDM\to\infty$ singularity 
at the GC (since $\psiGC\to 0$ there), and it helps to discriminate the large volume where the dark matter distribution is essentially flat.  

The definition of $\TSi$ is tied to the DM model and is not modified 
when evaluating the alternative. For the isotropic hypothesis 
$H_{\rm iso}$ we generate mock events drawn uniformly over the 
observable sky and evaluate the same Eq.~\eqref{eq:ts_def_1ev} 
against $\PDM$ and $\psiGC$. Because isotropically distributed events 
do not preferentially come from the GC, they populate regions with 
small $\PDM$ and large $\psiGC$, so the $H_{\rm iso}$ distribution of 
$\TStot$ peaks at significantly higher values than under $H_{\rm DM}$, 
where events cluster near the GC. The DM $p$-value is the fraction 
of simulated DM experiments with $\TStot$ at least as large as the observation,
\beq
\pval_{\rm DM}=P(\TStot\geq\TStot^{\rm obs}\mid H_{\rm DM})\,.
\label{eq:pvalue}
\eeq

\section{Results and discussion}
\label{sec:results}

We evaluate the framework in three settings of increasing information. First, we consider an idealized \emph{geometry-only} case where we assume $\Vis\equiv1$, i.e. the detector sees the whole sky at all times and the test depends only on the static angular geometry. Next, we consider the \emph{time and azimuthal-averaged} case, folding in the detector exposure through its daily average $ \langle\Vis(E,\hat n,t)\rangle_{t,\alpha}$. This is the level of information used in previous visibility-based discussions of the event~\cite{Aloisio:2025nts}. Finally, we take the \emph{time-resolved} case using the full $\PDM(t,\delta,\alpha)$, retaining the arrival time of each event. We present the analyses in these three settings to isolate how much discrimination power comes from angular geometry when including average exposure and timing alone. Unless stated otherwise we assume $\nu\bar\nu$ decay.

\begin{figure}[h!]
\centering
\includegraphics[width=1\linewidth]{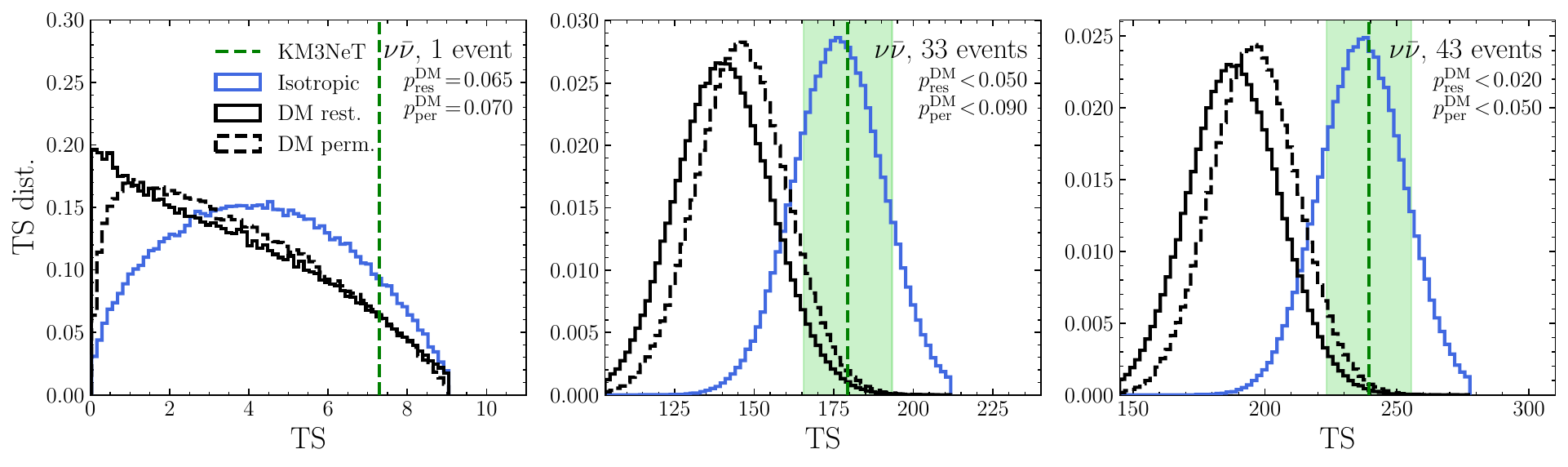}
\caption{Geometry-only scenario ($\Vis\equiv1$), $\nu\bar\nu$ channel. Green
dashed: TS of the observed event; solid (dashed) black: restrictive (permissive)
halo profile; blue: isotropic hypothesis. Panels (a)--(c): single event, and
combined samples of $33$ and $43$ events.}
\label{fig:ts_static}
\end{figure}

\subsection{Geometry only}

With visibility switched off, the single observed event yields $\pval_{\rm DM}=0.065$ (restrictive) and $0.07$ (permissive). We show the corresponding test-statistic distribution in Fig.~\ref{fig:ts_static}(a). We see that $\kmt$ (green dashed line) is disfavoured, but not decisively. Correspondingly, the DM and isotropic TS distributions overlap substantially. The halo plus extra-galactic signal, although peaked toward the GC, is still spatially broad, so a single direction carries limited discriminating power. Stacking the observed event with projected future detections drawn from the isotropic (astrophysical) hypothesis, we can determine how many future events would be required to distinguish between $H_{\rm DM}$ and $H_{\rm iso}$. We show when the cumulative $\pval_{\rm DM}$ falls below $0.05$ for the restrictive and permissive halo cases in Fig.~\ref{fig:ts_static}(b) and Fig.~\ref{fig:ts_static}(c) respectively. The median $1\sigma$ Asimov expectation is shown via the green band. We can see that the restrictive halo model can be excluded after $\sim33$ events, whereas for the permissive profile $\sim43$ events are required for the same level of discrimination. At $43$ events the restrictive profile is already excluded with $p<0.02$.

\begin{figure}[h!]
\centering
\includegraphics[width=1\linewidth]{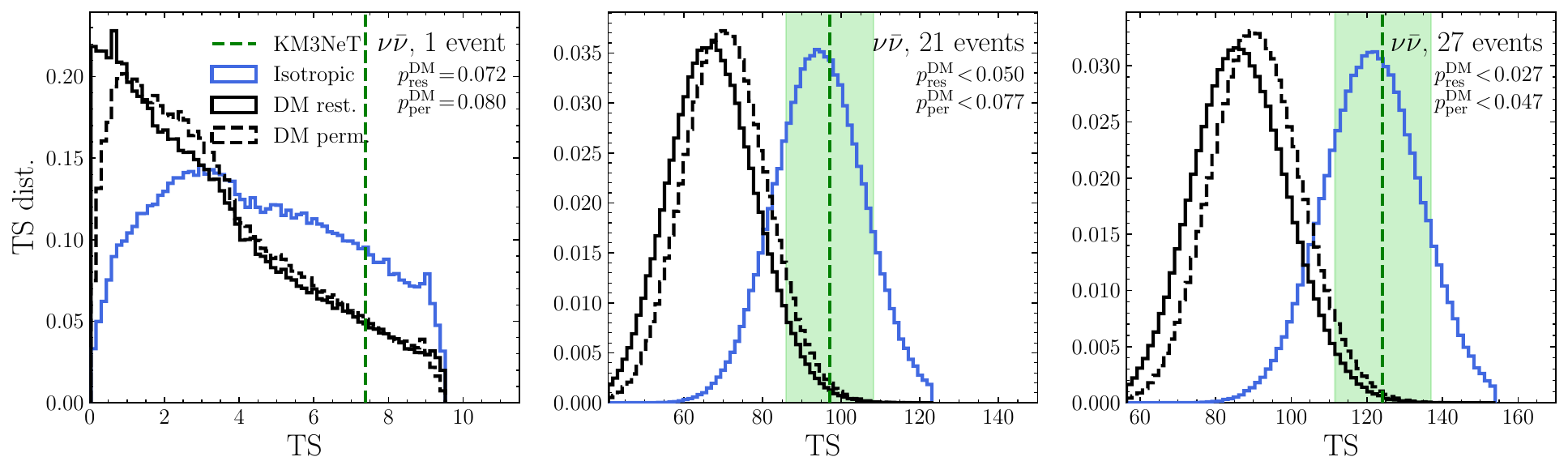}
\caption{Time-averaged-exposure scenario, $\nu\bar\nu$ channel: the visibility
enters only through its daily average, discarding
the arrival time. Otherwise as Fig.~\ref{fig:ts_static}. Panels (a)--(c): single
event, and combined samples of $22$ and $27$ events.}
\label{fig:ts_avg}
\end{figure}

\subsection{Time-averaged exposure}
Including the exposure only through its daily average already enhances the test
relative to the geometry-only case. In Fig.~\ref{fig:ts_avg} we can see that the cumulative 
$\pval_{\rm DM}$ now crosses $0.05$ after $\sim22$ events (restrictive) and $\sim27$ events (permissive), down from
$33$ and $43$. The improvement has a simple explanation. The average exposure is strongly 
declination-dependent (Sec.~\ref{sec:vis}; Fig.~\ref{fig:pdm_skymap}): small in the circumpolar cap, 
largest towards the south. Averaging $\PDM$ over time therefore suppresses part of the sky 
that an isotropic source would otherwise populate and re-weights the rest. This enhances the contrast between the GC-peaked DM
expectation and the isotropic hypothesis. At the single-event level shown in Fig.~\ref{fig:ts_avg}(a),
the average exposure leaves the picture essentially unchanged, $\pval_{\rm DM}=0.072$ (restrictive) and
$0.080$ (permissive), only marginally above the geometry-only values. One event from a fixed direction
barely samples the daily-averaged exposure, which is why the difference is only apparent through the
combined samples.

\begin{figure}[h!]
\centering
\includegraphics[width=1\linewidth]{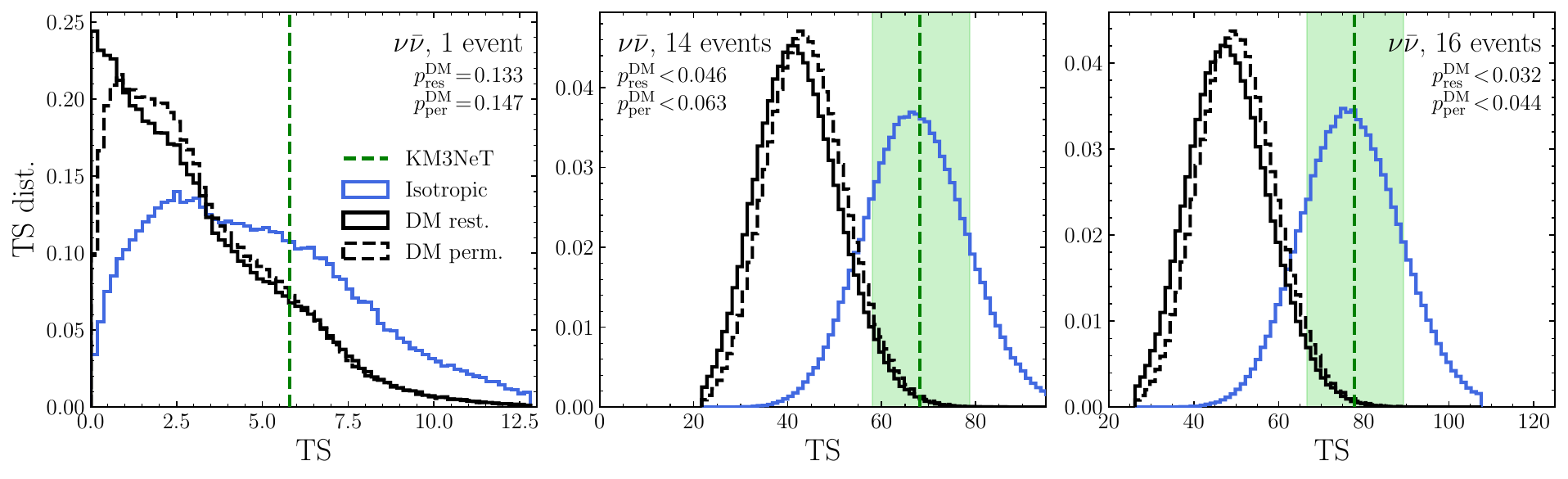}
\caption{Time-resolved scenario, $\nu\bar\nu$ channel: the full
$\PDM(t,\delta,\alpha)$ retains the arrival time of each event. Otherwise as
Fig.~\ref{fig:ts_static}. Panels (a)--(c): single event, and combined samples of
$14$ and $16$ events.}
\label{fig:ts_dynamic}
\end{figure}

\subsection{Time-resolved exposure: isolating the temporal gain}
The central result of this work is the \emph{further} reduction obtained when the exposure is resolved in time rather than averaged. We show in Fig.~\ref{fig:ts_dynamic} that replacing the time averaged PDF
by the full $\PDM(t,\delta,\alpha)$ lowers the threshold from $22$ to $14$ events (restrictive) and
from $27$ to $16$ (permissive), a reduction of $\sim33$--$38\%$ thanks to the arrival time alone. Of
the total reduction from the geometry-only to the time-resolved case ($33\!\to\!14$ and $43\!\to\!16$),
roughly $40\%$ comes purely from timing and is therefore comparable to the improvement from including
the average visibility function. The reason is that the daily average is blind to right ascension,
and it assigns the same exposure to every RA at a given declination, whereas the arrival time
$t_i$ selects \emph{which} RA strip was actually below the horizon at that instant. Two events sharing
a declination but recorded at different times probe different angular distances to the GC; averaging 
over the day erases this distinction. The size of the gain is set by the amplitude of the
daily modulation $\mathcal{R}^{\rm DM}(t)$ of Fig.~\ref{fig:rate_vis}: the deeper the modulation, the
more the timing helps.

In Fig.~\ref{fig:ts_dynamic} we show the results. At the single-event level
the the $\pval_{\rm DM}$ has shifted upward to $0.133$ (restrictive)
and $0.149$ (permissive). Including instantaneous visibility at the
observed arrival time makes the lone event less anomalous than the 
geometry-only analysis suggests, and somewhat more compatible with a
DM origin. 
Structurally, $\PDM$ is normalised over the
instantaneously visible sky and in the
geometry-only case ($\Vis\equiv1$) the normalisation integrates the entire GC-peaked flux, so an anti-centre direction carries a small relative
probability. Restricting to the thin near-horizon band from
the UHE event, removes most of the high-density region around the GC from the denominator, so the denominator falls faster than the numerator and the relative probability of an anti-centre
arrival increases. This alleviation of the anti-centre tension 
is generic to any up-going event. Its size for $\kmt$, 
however, is set by the arrival time. The GC and anti-GC at 
$t=0$, share the same visible band. An otherwise identical event 
arriving when the GC sits near the nadir, that deep in the opaque
Earth, would not enhance the p-value. The timing therefore makes the lone
event marginally more DM-compatible while, through the daily modulation, 
making a future sample far easier to discriminate.

\begin{figure}[h!]
\centering
\includegraphics[width=1\linewidth]{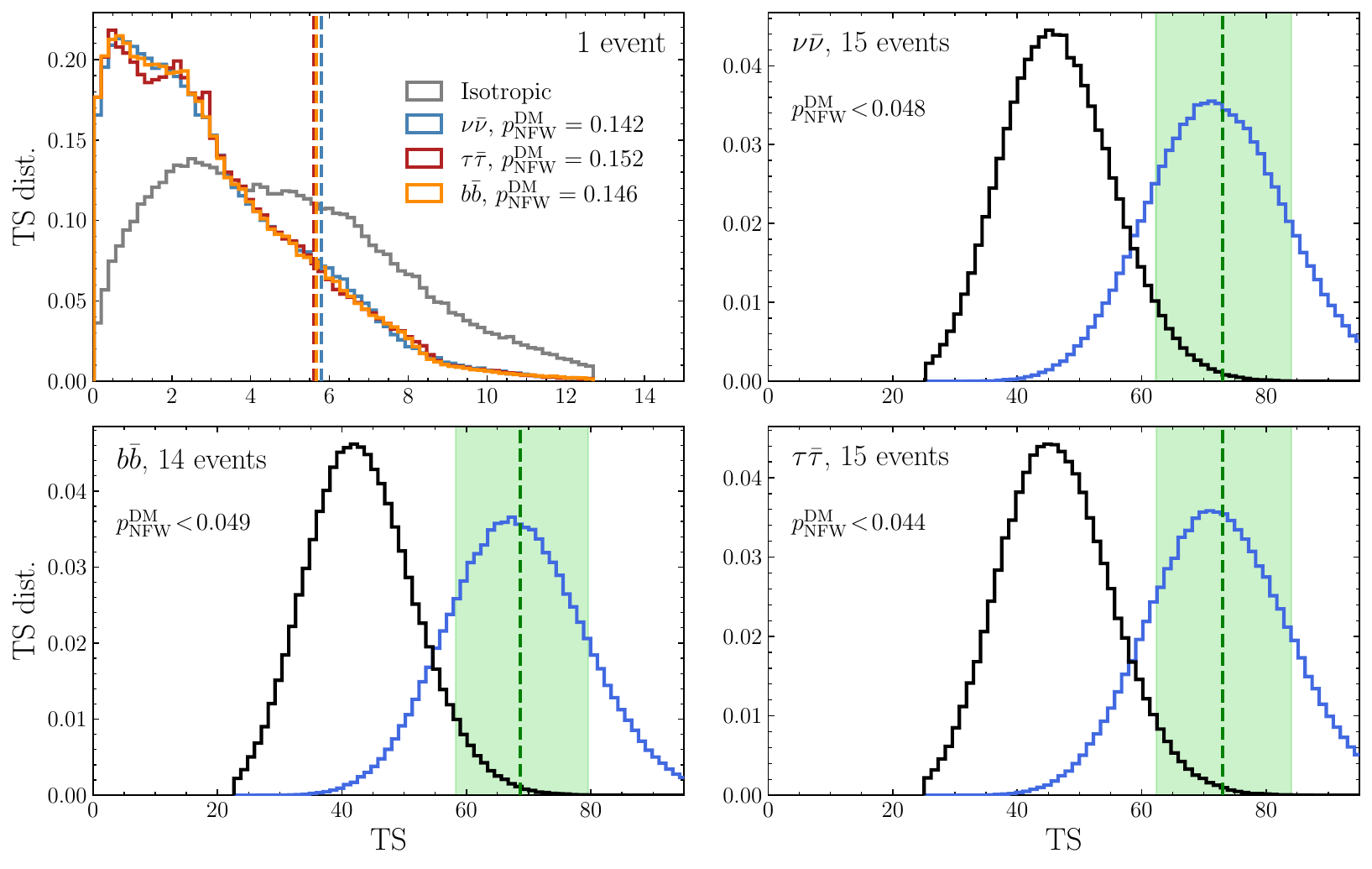}
\caption{Dependence on the DM decay channel, for the time-resolved scenario and
the NFW profile only. Grey: isotropic hypothesis; blue: $\bar\nu\nu$; red:
$\bar\tau\tau$; orange: $\bar b b$. Panel (a) shows the single observed event;
panels (b), (c), (d) the combined samples for $\bar\nu\nu$, $\bar b b$ and
$\bar\tau\tau$ respectively.}
\label{fig:ts_NFW}
\end{figure}

\subsection{Channel dependence}
The benchmark decay channels adopted by Ref.~\cite{KM3NeT:2026flu} differ mainly in spectral shape and end-point (Fig.~\ref{fig:fluxes_bmp}): the $\bar\nu\nu$ flux cuts off near $10^8$~GeV, the $\bar\tau\tau$ flux at $5\times10^8$~GeV, and the $\bar b b$ flux only at $\sim4\times10^{10}$~GeV. Due to the energy dependence of $\mathcal{V}$, such spectral differences alter the statistical gain from including timing information. Harder spectra reach higher energies, where the Earth is more opaque, the instantaneously visible band is narrower and the daily modulation $\mathcal{R}^{\rm DM}(t)$ has a larger amplitude, so the arrival time carries more information. Fig.~\ref{fig:ts_NFW} confirms this for the NFW profile. For the single event the $p$-values are nearly channel-independent ($\pval_{\rm DM}=0.142$, $0.152$, $0.147$ for $\bar\nu\nu$, $\bar\tau\tau$, $\bar b b$ [Fig.~\ref{fig:ts_NFW}(a)]), so no channel is specifically more or less compatible with $H_{\rm DM}$. However, the combined-event threshold for $p<0.05$ shrinks for the harder spectra: $15$ events for $\bar\nu\nu$ [Fig.~\ref{fig:ts_NFW}(b)] and $\bar\tau\tau$ [Fig.~\ref{fig:ts_NFW}(d)], while for $\bar b b$ only  $14$ events are required [Fig.~\ref{fig:ts_NFW}(c)]. The trend confirms that visibility/timing information, not just the angular geometry, drives the discrimination power for signal models. 

\subsection{Summary of exclusion reach}
The number of events needed to exclude a common DM origin at $\pval_{\rm DM}<0.05$ 
therefore falls from $\sim33$--$43$ (geometry only) to $\sim22$--$27$ (time-averaged) 
to $\sim14$--$16$ (time-resolved) as shown in Table~\ref{tab:pvalues}.

\begin{table}[h!]
\centering
\begin{tabular}{|c||c|c|c|}
\hline          
$\sum_i {\rm TS}_{\rm DM,i}$  & Restric. & Permis. & NFW \\
\hline    \hline
$\nu\bar{\nu}$ & 14& 16 &15 \\
\hline
$\tau\bar{\tau}$ &14 &15 & 15\\
\hline
$b\bar{b}$  &13 &15 & 14\\
\hline
\end{tabular}
\caption{Number of events required to exclude the DM hypothesis at
$\pval_{\rm DM}<0.05$, for each decay channel (rows) and MW DM density profile
(columns)}
\label{tab:pvalues}
\end{table}

Table~\ref{tab:pvalues} shows, for each decay channel and density profile, the number of events (including the observed one) needed to exclude the DM hypothesis at $p<0.05$. In all cases this number lies between $13$ and $16$. It decreases for the more centrally concentrated (restrictive) profile, which sharpens the contrast between the GC-peaked DM expectation and the isotropic one, and for the harder $b\bar b$ spectrum, whose higher characteristic energies make the Earth more opaque and thus enhance the visibility/time discrimination discussed above. We see that the uncertainties and associated with DM halo profile and final-state particles do not alter the required number of events much. Detecting $\sim 16$ events from around a similar energy scale can be seen then as a relevant experimental milestone for understanding the origin of $\kmt$.

\section{Conclusions}
In this work we have tried to assess whether the ultra-high-energy event $\kmt$ can be attributed to decaying dark matter in a statistically rigorous way. To do this we have developed a quantitative test built on two observables: the arrival \emph{direction} of the event, and its arrival \emph{time}. The central point of this work is that the two carry independent geometric information. Above PeV energies, the Earth is opaque, so KM3NeT/ARCA is sensitive only to a thin band near the horizon at a given instant. Throughout the day this band sweeps the sky. Therefore, the detection time tags a specific region of the sky. Whether the Galactic Centre is present in such a region holds significance.

To show the importance of including the arrival time information, we use three analyses to forecast how many events are required to exclude the standard DM hypothesis, $H_{\rm DM}$ in favour of a isotropic signal $H_{\rm iso}$. If we can ignore the detector's visibility and have only the signal hypotheses geometry, we require $\sim33$--$43$ events to exclude $H_{\rm DM}$. Including KM3NeT's visibility on the sky in a time-independent way lowers the required number of events for discrimination to $\sim22$--$27$. This is lowered by a further $40\%$ by including the fully time-dependent visibility, $\sim14$--$16$. This effect is strongest for harder decay spectra because the Earth is more opaque at higher energies, making the instantaneous visibility band narrower on the sky. 
This result is robust against DM halo uncertainties, since we have bracketed 
them with two limiting profiles spanning the range from centrally peaked 
to shallow-cored.

For the current situation where only one event has been detected, none of the three analyses can exclude the possibility that $\kmt$ originates from DM decay. The p-values are, $\pval_{\rm DM}=0.065$/$0.070$, $\pval_{\rm DM}=0.072$/$0.080$, and $\pval_{\rm DM}=0.133$/$0.149$ for the geometry only, time-averaged and time-resolved methods respectively. Interestingly, by including time arrival information, $\kmt$ becomes marginally \emph{more} DM-compatible, even as it makes a future \emph{sample} far easier to discriminate.

We argue that time information is decisive and potentially more important than spectral information when it comes to determining the origin of these neutrinos. A modest sample of order ten more UHE neutrinos would let KM3NeT confirm or rule out decaying DM as their common origin. This is a realistic target given the detector's projected exposure. More broadly, the method is generic to equatorial neutrino telescopes, where Earth opacity and detector rotation conspire to make event timing a probe of halo geometry. The same machinery applies to any anisotropic flux hypothesis, Galactic versus extragalactic, point-source versus diffuse. This work underlines the importance of retaining the time information generally and should be adopted by the next generation of equatorial telescopes such as TRIDENT~\cite{Ye:2023trident}, P-ONE~\cite{P-ONE:2020ljt} and Baikal-GVD~\cite{Avrorin:2011zzc}. Time-resolved directional analyses of this kind will become a standard tool for separating Galactic dark matter from astrophysical backgrounds.

\begin{acknowledgments}
JPP thanks the Universitat de Barcelona for its hospitality, which made it possible to discuss several crucial issues in person during the development of the paper. AC acknowledges the support of S. Ge, funded by the NSFC (Grant Nos. 12425506, 12375101, 12090060, and 12090064). JPP is supported by the National Natural Science Foundation of China (12425506 and 12375101). JS has been supported by the Spanish grant PID2022-126224NB-C21, ``Unit of Excellence Maria de Maeztu 2020-2023'' award to the ICC-UB CEX2019-000918-M and by the European Union’s Horizon 2020 research and innovation program under the Marie Skłodowska-Curie grants HORIZON-MSCA-2021-SE-01/101086085-ASYMMETRY and H2020-MSCA-ITN-2019/860881-HIDDeN.
\end{acknowledgments}


\appendix

\section{Detector visibility: geometry, Earth model and cross sections}
\label{app:visibility}

This appendix collects the ingredients of the visibility function
$\Vis(E,\hat n,t)$ of Eq.~\eqref{eq:visibility}: the chord geometry through the
Earth (App.~\ref{app:chord}), the effective target length set by the muon range
(App.~\ref{app:Leff}), the column depths in the PREM density model
(App.~\ref{app:columns}), the cross-section parametrisations (App.~\ref{app:xsec}),
and the mapping between arrival time and local zenith angle that gives $\Vis$ its
time dependence (App.~\ref{app:time}).

\subsection{Chord geometry}
\label{app:chord}
We model the detector as a single point at radius
$R_{\rm det}=R_\oplus-3.5\km=6367.5\km$ from the Earth's centre, with
$R_\oplus=6371\km$. A neutrino arriving from local zenith angle $\theta_z$
traverses a chord of total length $L_{\rm total}$, fixed by the law of cosines for
the triangle (Earth centre, detector, entry point),
\begin{equation}
L^2 + 2R_{\rm det}\cos\theta_z\,L + (R_{\rm det}^2 - R_\oplus^2) = 0
\;\Rightarrow\;
L_{\rm total} = -R_{\rm det}\cos\theta_z + \sqrt{R_\oplus^2 - R_{\rm det}^2\sin^2\theta_z}\,.
\label{eq:Ltotal}
\end{equation}
For up-going directions ($\theta_z>90^\circ$, $\cos\theta_z<0$) the chord length increases
as the trajectory approaches the nadir, reaching the full diameter
$\simeq2R_\oplus$ for a vertically up-going neutrino. Measuring the distance $l$
along the chord from the detector inward, the galactocentric radius at that point,
needed for the density integrals below, is
\begin{equation}
r(l)=\sqrt{R_{\rm det}^2 + l^2 + 2\,R_{\rm det}\,l\cos\theta_z}\,,
\label{eq:rofl}
\end{equation}
which decreases from $R_{\rm det}$ at the detector to its minimum at the chord
mid-point and back. Setting $r(L_{\rm total})=R_\oplus$ in Eq.~\eqref{eq:rofl}
recovers Eq.~\eqref{eq:Ltotal}.

\subsection{Effective length and muon range}
\label{app:Leff}
Only the portion of the chord within range of the instrumented volume can produce
a detectable lepton. We cap the interacting (``target'') segment at the
secondary-muon range,
\begin{equation}
L_{\rm eff} = \min\!\big(L_{\rm total},\,R_\mu(E)\big),
\qquad
R_\mu(E) = \frac{1}{B_{\rm water}}\ln\!\left(1 + \frac{B_{\rm water}\,E_\mu}{A_{\rm water}}\right),
\label{eq:Rmu}
\end{equation}
with $E_\mu = 0.75\,E_\nu$~\cite{Laha:2013lka}. The energy-loss parameters are
$A_{\rm water}=\alpha\,\rho_{\rm sw}$ and $B_{\rm water}=\beta\,\rho_{\rm sw}$, with
$\rho_{\rm sw}=1.04\,\mathrm{g\,cm^{-3}}$,
$\alpha\simeq2.0\times10^{-3}\,\mathrm{GeV\,cm^2/g}$ and
$\beta\simeq4.2\times10^{-6}\,\mathrm{cm^2/g}$ for
water~\cite{Chirkin:2004hz,Groom:2001kq}. At the energies of interest $R_\mu$
reaches the kilometre scale, so for most up-going directions
$L_{\rm eff}=R_\mu\ll L_{\rm total}$ and the chord splits cleanly into a short
target near the detector and a long attenuating screen.

\subsection{Column depths and Earth model}
\label{app:columns}
The target and screen segments are $[0,L_{\rm eff}]$ and
$[L_{\rm eff},L_{\rm total}]$ respectively, measured from the detector. Their
column depths are line integrals of the local density along the chord,
\begin{equation}
X_{\rm target} = 10^5\!\int_0^{L_{\rm eff}}\!\rho_{\rm PREM}\big(r(l)\big)\,dl,
\qquad
X_{\rm screen} = 10^5\!\int_{L_{\rm eff}}^{L_{\rm total}}\!\rho_{\rm PREM}\big(r(l)\big)\,dl
\quad[\mathrm{g\,cm^{-2}}],
\label{eq:columns}
\end{equation}
with $r(l)$ from Eq.~\eqref{eq:rofl}, the factor $10^5$ converting
$\mathrm{g\,cm^{-3}}\cdot\km$ to $\mathrm{g\,cm^{-2}}$, and $\rho_{\rm PREM}$ the
spherically symmetric density of the Preliminary Reference Earth
Model~\cite{Dziewonski:1981xy}. Because $X_{\rm screen}$ grows steeply for
trajectories that sample the dense inner core, the survival factor in
Eq.~\eqref{eq:visibility} suppresses all but the near-horizon directions, defining
the visible band discussed in Sec.~\ref{sec:vis}.

\subsection{Cross sections}
\label{app:xsec}
We adopt power-law fits to the UHE neutrino--nucleon cross
section~\cite{Arguelles:2025ewg},
\begin{equation}
\sigma_{\rm tot}(E)\approx 2.35\times10^{-32}
\left(\frac{E}{10\,\EeV}\right)^{0.363}\mathrm{cm}^2,
\qquad
\sigma_{\rm CC}(E)\approx 1.66\times10^{-32}
\left(\frac{E}{10\,\EeV}\right)^{0.363}\mathrm{cm}^2,
\label{eq:sigma_app}
\end{equation}
which enter the survival ($\sigma_{\rm tot}$) and interaction ($\sigma_{\rm CC}$)
factors of Eq.~\eqref{eq:visibility} respectively. The $E^{0.363}$ growth 
underlies the channel dependence of the discrimination power found in 
Sec.~\ref{sec:results}.

\subsection{From arrival time to zenith angle}
\label{app:time}
The time dependence of $\Vis$ enters entirely through the local zenith angle of a
fixed celestial direction. For a detector at geographic latitude $\phi$, a source
at declination $\delta$ and right ascension $\alpha$ has zenith angle
\begin{equation}
\cos\theta_z(t) = \sin\phi\,\sin\delta + \cos\phi\,\cos\delta\,\cos H(t),
\qquad
H(t) = \mathrm{LST}(t) - \alpha,
\label{eq:zenith}
\end{equation}
where $H$ is the hour angle and $\mathrm{LST}(t)$ the local  time, which
advances by $2\pi$ per day ($\simeq 23^{\rm h}56^{\rm m}$). Thus at fixed
$(\delta,\alpha)$ the chord geometry and through
Eqs.~\eqref{eq:Ltotal}--\eqref{eq:columns} the whole of $\Vis$, oscillates once
per day as $H$ runs over $2\pi$. Equation~\eqref{eq:zenith} is the precise
statement that the arrival time labels which right-ascension strip lay within the
visible band, and is what promotes $t_i$ to an independent observable in the test
statistic of Sec.~\ref{sec:ts}.

\section{Robustness of the test-statistic definition}
\label{app:ts_consistency}

This appendix supports the claim of Sec.~\ref{sec:ts} that the
$\psiGC$ weight in Eq.~\eqref{eq:ts_def_1ev} affects neither the
single-event $p$-value nor the qualitative behaviour of the
combined test. We compare with the unweighted statistic
\begin{equation}
\TSi^{\rm std} = -\log\PDM(t_i,\delta_i,\alpha_i)\,,
\label{eq:ts_std}
\end{equation}
i.e. Eq.~\eqref{eq:ts_def_1ev} with $\psiGC\to 1$. We have also
checked the more general family $-\log\PDM\times\psiGC^{\,n}$ with
$n=\tfrac12,1,2$ and find no improvement over $n=1$.

For the observed event the two definitions yield nearly identical
$p$-values, as expected for a global angular reweighting that
mostly rescales the TS axis. The difference is presentational:
under $\TSi^{\rm std}$ the anti-centre event lands close to the
mode of the $H_{\rm DM}$ histogram, suggesting visually—though
not statistically—that anti-centre arrivals are typical of a DM
origin; the $\psiGC$ factor pushes it into the tail where it
physically belongs.

For combined samples the unweighted statistic is the marginally
stronger discriminator. Table~\ref{tab:pvalues_TS} gives the
thresholds at $\pval_{\rm DM}<0.05$ for $\TSi^{\rm std}$: in every
channel and profile, exclusion is reached one or two events
earlier than with the $\psiGC$-weighted statistic
(Table~\ref{tab:pvalues}). The angular weight adds no
information once several events are combined, and the
single-event visual artefact also disappears as the $H_{\rm DM}$
and $H_{\rm iso}$ distributions separate cleanly with $N$. The
reported thresholds in the main text are therefore mildly
conservative, and none of the conclusions of Sec.~\ref{sec:results}
depend on the choice of weighting.

\begin{table}[h!]
\centering
\begin{tabular}{|c||c|c|c|}
\hline
$\sum_i \TSi^{\rm std}$ & Restric. & Permis. & NFW \\
\hline\hline
$\nu\bar{\nu}$   & 12 & 14 & 13 \\
\hline
$\tau\bar{\tau}$ & 12 & 13 & 13 \\
\hline
$b\bar{b}$       & 11 & 13 & 12 \\
\hline
\end{tabular}
\caption{As Table~\ref{tab:pvalues}, but for the unweighted
statistic $\TSi^{\rm std}=-\log\PDM$ of Eq.~\eqref{eq:ts_std}.
Thresholds are uniformly one--two events lower than for the
$\psiGC$-weighted statistic.}
\label{tab:pvalues_TS}
\end{table}

\bibliographystyle{JHEP}
\bibliography{refs}

\end{document}